\documentclass[12pt]{article}
\usepackage{graphics,color}
\begin{document}
\thispagestyle{empty}
\noindent\
\begin{center}
\large \bf Oscillating Neutrinos and Majorana Neutrino Masses
\end{center}
\hfill
 \vspace*{1cm}
\noindent
\begin{center}
{\bf Harald Fritzsch}\\
Department f\"ur Physik\\ 
Universit\"at M\"unchen\\
Theresienstra{\ss}e 37\\
Germany
\vspace*{0.5cm}
\end{center}

\begin{abstract}

We discuss the mass matrices with texture zeros for the quarks and leptons. The flavor mixing angles for the quarks are functions of the quark masses and can be calculated. The results agree with the experimental data.\\
The texture zero mass matrices for the leptons and the see-saw mechanism are used to derive relations between the matrix elements of the lepton mixing matrix and the ratios of the neutrino masses. Using the measured neutrino mass differences, the neutrino masses can be calculated.\\ 
The neutrinoless double beta decay is discussed. The effective Majorana neutrino mass, describing the neutrinoless double beta decay, can be calculated -  it is about 4.6 meV. The present experimental limit is at least twenty times larger. 

\end{abstract}

\newpage

The flavor mixing of the quarks is described by the CKM matrix:
\begin{eqnarray}
V^{}_{CKM}= \left( \matrix{ V^{}_{ud}
& V^{}_{us} & V^{}_{ub} \cr
V^{}_{cd} &
V^{}_{cs} &
V^{}_{cb} \cr V^{}_{td} & V^{}_{ts} & V^{}_{tb} \cr} \right). 
\end{eqnarray} 

The absolute values of the nine matrix elements have been measured in many experiments:
\begin{eqnarray}
	V^{}_{CKM}=> \left( \matrix{ 0.974 & 0.224 & 0.004 \cr 0.218 & 0.997 & 0.042 \cr
0.008 & 0.040 & 1.019 \cr} \right) \;.
\end{eqnarray}

There are several ways to describe the CKM-matrix in terms of three angles and one phase parameter. I prefer the parametrization, which Z. Xing and I introduced years ago (ref.(1)), given by the angles $\theta_u$, $\theta_d$, $\theta$ and a phase parameter $\phi$, which describes CP violation:
\begin{eqnarray}
V^{}_{CKM}= \left( \matrix{ c^{}_{u}
& s^{}_{u} & 0 \cr
-s^{}_{u} &
c^{}_{u} &
0 \cr 0 & 0 & 1 \cr} \right)\times \left( \matrix{ e^{-i\phi}
& 0 & 0 \cr
0 &
c &
s \cr 0 & -s & c \cr} \right) \times \left( \matrix{ c^{}_{d}
& - s^{}_{d} & 0 \cr
s^{}_{d} &
c^{}_{d} &
0 \cr 0 & 0 & 1 \cr} \right).
\end{eqnarray}
Here we used the short notation: $c^{}_{u,d} \sim \cos\theta^{}_{u,d}$, $s^{~}_{u,d}
\sim \sin\theta^{}_{u,d}$, $c \sim \cos\theta$ and $s \sim
\sin\theta$.\\

 Relations between the quark masses and the mixing angles can be derived, if the quark mass matrices have "texture zeros", as shown by S. Weinberg and me in 1977 (ref.(2)). Here are the mass matrices with "texture zeros" for six quarks:

\begin{eqnarray}
M= \left( \matrix{ 0 & A & 0 \cr A^* & 0 & B \cr
0 & B^* & C \cr} \right) \;.
\end{eqnarray}

We can now calculate the angles $\theta_u$ and   
$\theta_d$ as functions of the mass eigenvalues: 

\begin{equation}
\theta^{}_d \simeq
\sqrt{m^{}_d/m^{}_s},\hspace*{1cm}
\theta^{}_u \simeq
\sqrt{m^{}_u/m^{}_c}.
\end{equation}
\\
For the masses of the quarks we use the the values, given by PDG (ref.(3)): 

\begin{eqnarray}
m_u (2~GeV) \simeq 3~MeV , \nonumber \\
m_d (2~GeV) \simeq 5~MeV , \nonumber \\
  m_s (2~GeV) \simeq 95~MeV , \nonumber \\
   m_c (2~GeV)\simeq 970 ~MeV.
\end{eqnarray}

Using these masses for the quarks, we find for the mixing angles: 

\begin{equation}
\theta^{}_d \simeq (12.9\pm 0.4)^\circ,\hspace*{1cm} \theta^{}_u \simeq (3.2\pm 0.7)^\circ. 
\end{equation}

The experimental values agree with the theoretical results:
\begin{equation}
\theta^{}_d \simeq (11.7\pm 2.6)^\circ,\hspace*{1cm} \theta^{}_u \simeq (5.3\pm 1.4)^\circ.
\end{equation}

In a similar way one can calculate the third angle $\theta$. We find $\theta \simeq 2.5^\circ$, in agreement with experiment.\\
 
In the Standard Theory of particle physics the neutrinos do not have a mass. But a mass term can be introduced analogous to the mass term for the electrons. Nevertheless the masses of the neutrinos must be 
very small, much smaller than the mass of the electron. According to the limit from cosmology the sum of the neutrino masses must be less than 0.12 eV.\\

If the neutrinos have a small mass and if they are superpositions of mass eigenstates, there would be also a flavor mixing of the leptons. An electron neutrino, emitted from a nucleus, can turn into a muon neutrino after travelling a certain distance. Afterwards it would again become an electron neutrino, etc. Thus neutrinos oscillate (see ref.(4)).\\

The flavor mixing of the leptons is described by a unitary 3x3-matrix, which is similar to the CKM-matrix for the quarks:

\begin{eqnarray}
U= \left( \matrix{ U^{}_{e1}
& U^{}_{e2} & U^{}_{e3} \cr
U^{}_{\mu1} &
U^{}_{\mu2} &
U^{}_{\mu3} \cr U^{}_{\tau1} & U^{}_{\tau2} & U^{}_{\tau3} \cr} \right). 
\end{eqnarray} 

This matrix can be described by three angles and a phase parameter. Here we use the standard parametrization of the mixing matrix, given by the three angles $\theta^{}_{12}$, $\theta^{}_{13}$ and $\theta^{}_{23}$.\\

In the nuclear fusion on the sun many electron neutrinos are produced. In 1963 John Bahcall calculated the flux of the solar neutrinos. He concluded that this flux could be measured by experiments. Raymond Davis (Brookhaven National Laboratory) prepared such an experiment. It was placed in the Homestake Gold Mine in Lead, South Dakota and took data from 1970 until 1994. One observed only about 1/3 of the flux, calculated by Bahcall. Today we know that the reduction of the solar neutrino flux is due to neutrino oscillations.\\

In the Japanese Alps, near the small village “Kamioka”, a big detector was built in 1982. It is located about 1000 m underground. This detector “Kamiokande” was built in order to find the hypothetical decay of a proton. Thus far no proton decay has been observed, but the detector can also be used to study neutrinos, in particular the atmospheric neutrinos, produced by the decay of pions in the upper atmosphere.\\

In 1996 the new detector “Superkamiokande” started to investigate these neutrinos. This detector consists of a water tank, containing 50 000 liters of purified water, surrounded by about 11 000 photo multipliers. With this detector one could measure the flux of the neutrinos. The flux of neutrinos, coming from the atmosphere above Kamioka, was as high as expected, but the flux of the neutrinos, coming from the other side of the earth, was only about 50\% of the expected rate. Afterwards a neutrino beam, sent from the KEK laboratory near Tsukuba towards Kamioka, was investigated. Again the flux of muon neutrinos was less than expected.\\ 

Oscillations between the muon neutrinos and the tau neutrinos could explain the observed reduction of the flux. These oscillations are described by the angle $\theta^{}_{23}$. According to the experiments this angle is very large:
\begin{eqnarray}
40.3^\circ \leq \theta_{23} \leq 52.4^\circ.
\end{eqnarray} 

In Canada a neutrino detector was built near Sudbury (Ontario), the Sudbury Neutrino Observatory (SNO). With this detector one could observe the solar neutrinos. If a solar neutrino collides with a deuteron, two protons and an electron are emitted. This process can be observed. Furthermore it was possible to observe the neutral current interaction of the neutrinos. If a solar neutrino collides with a deuteron, it splits up into a proton and a neutron. Also this reaction can be observed.\\

The neutral current interaction is not affected by oscillations, since all neutrinos have the same neutral current interaction. However oscillations can be observed for the charged current interaction. An electron neutrino, which becomes a muon neutrino, will not produce an electron after colliding with a nucleus. \\

By comparing the interaction rates for the neutral and for the charged current interactions one has observed the 
oscillations of the solar neutrinos. The mixing angle $\theta^{}_{12}$ was measured:
\begin{eqnarray}
31.6^\circ \leq \theta_{12} \leq 36.3^\circ.
\end{eqnarray} 

Nuclear reactors emit electron antineutrinos. These neutrinos have been investigated at a few nuclear reactors, in particular at the four Daya Bay reactors in China. Here neutrino oscillations have been observed, and one could measure the mixing angle $\theta^{}_{13}$:
\begin{eqnarray}
8.2^\circ \leq \theta_{13} \leq 9.0^\circ.
\end{eqnarray} 

Also the two small mass differences between the three neutrinos have been measured:
\begin{eqnarray}
\Delta m^2_{21} = (7.53 \pm 0.06)\times 10^{-5} ~{\rm eV}^2 , \nonumber \\
| \Delta m^2_{32}|  \simeq | \Delta m^2_{31}|  =  (2.44 \pm 0.06) \times 10^{-3} ~{\rm eV}^2.
\end{eqnarray}

The neutrino mass differences are very small, and the question arises, if the neutrino masses are different from the Dirac masses of the charged leptons. Since the neutrinos are neutral, the neutrino masses might be Majorana masses.\\

The smallness of the neutrino masses can be understood by the "seesaw"-mechanism (ref. 5). The mass matrix of the neutrinos is a matrix with one "texture zero" in the (1,1)-position. The two off-diagonal terms are given by the Dirac mass term D - a large Majorana mass term is in the (2,2)-position:
\begin{eqnarray}
M_\nu = \left( \matrix{ 0 & D \cr D & M \cr
} \right)\;.
\end{eqnarray}
After diagonalization one obtains a large Majorana mass M and a small neutrino mass: 
\begin{equation}
  {m^{}_\nu}\simeq D^{2}/M. 
\end{equation}
Now we assume that the Dirac mass matrices of the leptons also have four texture zeros:
\begin{eqnarray}
M^{}_D= \left( \matrix{ 0 & A & 0 \cr A^* & 0 & C \cr
0 & C^* & D \cr} \right) \;.
\end{eqnarray}

In the seesaw formula we replace the Dirac mass by the texture zero mass matrix $M^{}_D$ and the Majorana mass by a Majorana mass matrix $M^{}_R$:
\begin{eqnarray}
M_\nu = M^T_{\rm D} M^{-1}_{\rm R} M^{}_D.
\end{eqnarray}

Since the Majorana masses are much larger than the masses of the leptons and quarks, we assume, that the Majorana mass matrix is proportional to the unit matrix.
In this case the mixing angles are functions of the ratios of the charged lepton masses and of the neutrino masses (see also ref.(6)).\\

But the mass ratios of the charged leptons are very small and cannot give large mixing angles. These angles must be related to large ratios of the neutrino masses (ref.(7,8,9,10)).  
In a good approximation one can neglect the mass ratios of the charged leptons. We calculate the matrix elements of the mixing matrix, in the particular these matrix elements: 

\begin{eqnarray}
|U_{e2}|  \cong \left(\frac{m_1}{m_2}\right)^{1/4}  , \nonumber \\
|U_{\mu3}|  \cong \left(\frac{m_2}{m_3}\right)^{1/4}, \nonumber \\
|U_{e3}|  \cong \left(\frac{m_2}{m_3}\right)^{1/2} \left(\frac{m_1}{m_3}\right)^{1/4}\; .
\end{eqnarray}\\ 

The first and the second relation imply that there is a normal hierarchy for the neutrino masses: The mass of the first neutrino is smaller than the mass of the second neutrino mass, and this mass
 is smaller than the mass of the third neutrino. We use these two relations, the experimental values of the mixing angles and the relations (13) to 
determine the three neutrino masses. We obtain: 
\begin{eqnarray}
0.6~meV <{m^{}_1}<1.0~meV , \nonumber \\
8.6~meV <{m^{}_2}<8.8~meV , \nonumber \\
50~meV <{m^{}_3}<54~meV.
\end{eqnarray}
\\
The third equation in (18) implies a relation between $\theta_{13}$ and the neutrino mass ratios. If we take the central values of the three neutrino masses, we find $\theta_{13}$=$8.3^\circ$, in agreement 
with relation (12). \\ 

One expects that the Dirac masses of the neutrinos are similar to the corresponding charged lepton masses. For example, let us consider the tau lepton and its neutrino. If the Dirac mass of the tau neutrino 
is given by the tau lepton mass, we obtain for the heavy Majorana mass M:
\begin{equation}
M \simeq 6.3 \times 10^{10}~ GeV. 
\end{equation}
Thus the Majorana mass must be very large, almost as large as the mass, used in the Grand Unified Theories of quarks and leptons.\\

The only way to test the nature of the neutrino masses is to study the neutrinoless double beta decay, which violates lepton number conservation. Two neutrons inside an atomic nucleus decay by emitting two electrons and two neutrinos. The two Majorana neutrinos annihilate - only two electrons are emitted. The annihilation rate is a function of the Majorana mass of the neutrino. It would be zero, if the neutrino mass vanishes. \\

If neutrinos mix, all three neutrino masses will contribute to the decay rate. Their contributions are given by the masses of the neutrinos and by the mixing angles. The decay rate depends on the effective neutrino mass, which is a function of the three neutrino masses and the associated mixing angles: 
\begin{equation}
\widetilde{m}= | \sum^3_{i=1} m^{}_i U^2_{e i}| .
\end{equation}
Using the neutrino masses, given in relation (19), and the observed mixing angles, we can calculate the effective neutrino mass, relevant for the neutrinoless double beta decay (see also ref.(11)): 
\begin{equation}
  \widetilde{m}\simeq 4.6~meV.\textbf{}
\end{equation}
All three neutrinos contribute to this effective mass - the second neutrino about $55\%$, the third neutrino about $27\%$ and the first neutrino about $18\%$.\\ 

In various experiments one has searched for the neutrinoless double beta decay, for example in the decay of tellurium. Thus far the decay has not been observed. Here is the present limit for the effective neutrino mass, given by the Cuore and the Gerda experiments in the Gran Sasso Laboratory: 
 \begin{equation}
  \widetilde{m} < 110~meV.
\end{equation} 
This limit is at least twenty times larger than the value, which we have calculated.\\

Soon there will be a new experiment in the Gran Sass Laboratory, the experiment LEGEND. With this experiment, starting about 2025, it will be possible to reduce the limit for the effective Majorana neutrino mass to about 1~meV.\\ 

Since we calculated this mass to about 5~meV, we expect that the neutrinoless double beta decay is observed by the LEGEND experiment. If the effective Majorana mass is measured to about 4.6 meV, the Majorana neutrino masses should be about equal to the neutrino masses, which we calculated.

\end{document}